\documentclass[twocolumn,showpacs,preprintnumbers,amsmath,amssymb]{revtex4}
\usepackage{graphicx}
\usepackage{dcolumn}
\usepackage{bm}

\begin{document}

\title{Flow Equation Calculation of Transient and Steady State Currents in the Anderson Impurity Model}

\author{P. Wang}
\email{pei.wang@physik.lmu.de}
\author{S. Kehrein}
\affiliation{
Physics Department, Arnold Sommerfeld Center for Theoretical Physics and Center for NanoScience, \\
Ludwig-Maximilians-Universit\"at, Theresienstrasse 37, 80333 Munich, Germany
}

\date{\today}

\begin{abstract}
Transient and steady state currents through dc-biased quantum impurity models beyond the linear response
regime are of considerable interest, both from an experimental and a theoretical point of view. 
Here we present a new analytical approach for the calculation of such currents based on the
flow equation method (method of infinitesimal unitary transformations). Specifically, we 
analyze the Anderson impurity model in its mixed valence regime where the coupling to the
leads is switched on suddenly at time $t=0$. We observe the real time buildup of the current
until it reaches its steady state limit.
\end{abstract}

\pacs{02.70.-c, 72.15.Qm}

\maketitle

\section{\label{sec:level1} Introduction}

Transport properties of quantum devices beyond the linear response regime have generated a lot of interest
in the past decade. Experimentally, this is due to the recent advances in nanotechnology that permit
to apply large electrical fields in low dimensional electronic structures.
Theoretically, transport beyond the linear response regime is interesting
since it explores genuine non-equilibrium quantum many-body phenomena. A particularly
well-studied case, both experimentally and theoretically, are quantum dots in the
Coulomb blockade regime that display Kondo physics \cite{Gordon98,Cronenwett98,klitzing_1998}: 
here the shot noise generated 
by the steady state current serves as a source of decoherence that suppresses the
Kondo quasiparticle resonance for sufficiently large voltage bias \cite{Rosch2001}, thereby reducing 
the differential conductance \cite{Wiel00}. 

However, the interplay of correlation physics and non-equilibrium is difficult
to address theoretically, in spite of considerable effort in recent years. 
New numerical methods have been developed like
the scattering state numerical renormalization group \cite{AndersPRL101}, 
Monte Carlo methods \cite{Schiro2009,Millis2009},
the time-dependent density matrix renormalization group \cite{Boulat2008,HeidrichMeissner2009,Feiguin2008,Silva08}
and other real time methods \cite{Egger2008,FabianNewJPhys}.
Analytical approaches are perturbative Keldysh calculations \cite{ueda}, 
extensions of the renormalization group 
\cite{KehreinPRL,Fritsch,schoellerRG,jakobs:150603,Korb2007,Schoeller2009a,Borda2007,mitra:085342,segal:195316,Rosch2001,Rosch2005,Paaske2004,Schoeller2010,Karrasch2009}, 
generalizations of NCA (non-crossing approximation) to non-equilibrium \cite{nordlander,plihal,nordlander2005}, 
$1/N$-expansions \cite{Aditi2009}, 
Gutzwiller methods \cite{Spataru2009} and various approaches 
builing on integrability \cite{Schiller2000,Konik2001,Mehta06,boulat2008}. 
Since all of these methods have their respective limited domain of 
applicability, there is still a need for new theoretical methods. 

In the past few years the flow equation method (method of infinitesimal unitary
transformations) \cite{wegner1994,kehreinbook}
was used for a number of non-equilibrium quantum many-body problems
like interaction quenches \cite{Moeckelrev,hacklferroKondo,sabio}
and dc-transport beyond the linear response regime \cite{KehreinPRL,Fritsch,Fritsch2010}. 
In particular, for the Kondo model numerous quantities like spin susceptiblity,
magnetization and T-matrix have been calculated for large voltage bias in the steady 
state \cite{Fritsch,Fritsch2010}. In addition,
the flow equation method is particularly suitable for calculating the real time 
evolution of non-equilibrium problems \cite{hackl}. 
Therefore it offers the possibility to study
the transient time-dependent buildup of a quantity until it reaches its steady
state value, see for example the calculation of the magnetization dynamics in the
ferromagnetic Kondo model \cite{hacklferroKondo}. 
This defines the question investigated in this paper: Using the 
flow equation method, we calculate
the time-dependent buildup of the electrical current through an Anderson impurity
model when the coupling between the leads and the quantum dot is suddenly switched on 
at time $t=0$. Thereby we develop a new analytical method for calculating transport 
properties of interacting quantum systems beyond the linear response regime, both for
transient and steady state behavior.

The model of a single level quantum dot coupled to two leads is described by the Anderson impurity Hamiltonian:
\begin{eqnarray} \label{model}\nonumber
H &=& \sum_{k\sigma \alpha} \epsilon_k c^\dag_{k\alpha \sigma} c_{k\alpha \sigma} + \epsilon_d \sum_\sigma d^\dag_{\sigma} d_\sigma \\ && +  \sum_{k \alpha \sigma} 
\frac{V}{\sqrt{2}} (c^\dag_{k \alpha \sigma} d_\sigma + h.c.) 
 + U\, d^\dagger_\uparrow d_\uparrow d^\dagger_\downarrow d_\downarrow \ ,
\end{eqnarray}
$k$ denotes the wave vector, $\sigma=\uparrow,\downarrow$ the electron spin and $\alpha=L,R$ the left and right lead.  
For time $t<0$ both leads are in equilibrium at different chemical potentials $\mu_L$
and $\mu_R$. The hybridzation $V$ between leads and the dot is then switched on at time $t=0$
and we are interested in the current $I(t)$ as a function of time. For simplicity we restrict
ourselves to symmetric coupling to the leads, although the calculation can be generalized in
a straightforward way.

An explicit expression for $I(t)$ is achieved via the forward-backward technique of the flow equation
method \cite{hackl}: The current operator is expressed in the diagonal basis of the Hamiltonian (\ref{model}), 
where its time evolution can be worked out easily. Then the time-evolved operator is transformed
back into the original basis, where the initial condition of non-interacting Fermi gases with
different chemical potentials is given. This yields the final answer with an explicit expression for the current as
a function of time. Approximations enter during the diagonalization step of the Hamiltonian, which
limits our calculation to weak and intermediate interaction~$U$. However, the voltage bias can be
large (beyond the linear response regime) and the real time evolution followed into the asymptotic
steady state limit without any difficulties. 
 
\section{Transformation of the Hamiltonian}

We employ a symmetric/antisymmetric basis $c_{k\pm\sigma}=\frac{1}{\sqrt{2}}(c_{kL\sigma}\pm c_{kR\sigma})$
and re-express the Hamiltonian as
\begin{eqnarray} \nonumber
H &=& \sum_{k\sigma} \epsilon_k (c^\dag_{k + \sigma} c_{k + \sigma}+ c^\dag_{k - \sigma} c_{k - \sigma})+ \epsilon_d \sum_\sigma d^\dag_\sigma d_\sigma \\
&& +
\sum_{k\sigma} V (c^\dag_{k + \sigma} d_\sigma + h.c.) +U\, d^\dagger_\uparrow d_\uparrow d^\dagger_\downarrow d_\downarrow \ .
\end{eqnarray}
Notice that only the symmetric combination of lead operators couples to the impurity orbital, which
plays an important role in the solution later.

In order to work out the flow equation solution for the current, it turns out to be convenient to use a finite
system with a discrete level spacing. The thermodynamic limit will then be taken at the very end when
the current is evaluated. We take a constant level spacing $\Delta$ corresponding to a constant and equal density of states
$\rho=1/\Delta$
in both leads. The symmetric non-interacting terms in the Hamiltonian can then be diagonalized~\cite{Moeckel}
\begin{eqnarray}\nonumber
\sum_{k\sigma} \epsilon_k c^\dag_{k + \sigma} c_{k + \sigma}+
\sum_{k\sigma} V (c^\dag_{k + \sigma} d_\sigma + h.c.)=\sum_{s\sigma} \epsilon_s c^\dag_{s \sigma} c_{s \sigma}. \\
\end{eqnarray}
by defining the {\it pre-diagonalized} basis
\begin{eqnarray}
c_{s \sigma} = \sum_k \frac{V}{\epsilon_s - \epsilon_k} B_s c_{k + \sigma} + B_s d_\sigma,
\end{eqnarray}
with the transformation coefficient $B_s = \frac{V}{\sqrt{\epsilon_s^2 + \Gamma ^ 2}}$ and the linewidth
$\Gamma=\rho \pi V^2$.
The inverse tranformation is $d_\sigma = \sum_s B_s c_{s \sigma}$ and through this
the interaction term can also be expressed in the pre-diagonalized basis:
\begin{equation}
U\,n_\uparrow n_\downarrow= \sum_{s_1's_1s_2's_2} U B_{s_1'}B_{s_1}B_{s_2'}B_{s_2} c^\dag_{s_1' \uparrow} c_{s_1 \uparrow} c^\dag_{s_2' \downarrow} c_{s_2 \downarrow} \ .
\end{equation}

In the sequel we will work with normal-ordered expressions.
In this model we define normal ordering with respect to the non-interacting ground state
in equilibrium, which is also the initial state at time $t=0$.
The chemical potentials of the left and right lead are $\mu_L$ and $\mu_R$, respectively, 
and $V_{sd}=\mu_R-\mu_L$ denotes the voltage bias.
Strictly speaking, the Fermi function in the pre-diagonalized basis has no sharp edge
even at zero temperature due to hybridization. But this effect vanishes in the
thermodynamic limit and we can use
\begin{eqnarray}\nonumber
n_s &=& \langle c_{s \sigma}^\dag c_{s \sigma} \rangle_0
\\ &= & \frac{1}{2} (f_L(\epsilon_s)+f_R(\epsilon_s)) \label{nsdef}
\end{eqnarray}
with the usual Fermi function 
\begin{equation}
f_\alpha(\epsilon)=\frac{1}{1+e^{\beta(\epsilon-\mu_\alpha)}}
\end{equation}
In this paper we will generally work at zero temperature ($\beta=\infty$), the generalization
to non-zero temperature is straightforward.

The starting point of our analysis is the following Hamiltonian
\begin{eqnarray} \nonumber
H &=& \sum_k \epsilon_k c^\dag_{k-\sigma}c_{k-\sigma}+ \sum_{s\sigma} \epsilon_s c^\dag_{s \sigma} c_{s \sigma} \\
 && + \sum_{s_1's_1s_2's_2} U B_{s_1'}B_{s_1}B_{s_2'}B_{s_2} :c^\dag_{s_1' \uparrow} c_{s_1 \uparrow} c^\dag_{s_2' \downarrow} c_{s_2 \downarrow}: \ ,
\label{model2}
\end{eqnarray}
which corresponds to (\ref{model}) with a single-particle energy
$\epsilon_d = -\frac{U}{2} \sum_s B^2_s (f_L(\epsilon_s)+f_R(\epsilon_s))$. 
Notice that the energy of the impurity level is then related to the
lead chemical potentials, i.e. at zero temperature by
\begin{eqnarray}\nonumber\label{para_eq}
\epsilon_d - \mu &=& -\frac{U}{2} - \frac{U}{2\pi} [ \arctan( \mu-\frac{V_{sd}}{2}) \\&& +\arctan (\mu+\frac{V_{sd}}{2}) ] -\mu,
\end{eqnarray}
where $\mu=\frac{\mu_L+\mu_R}{2}$. The natural parameters in an experiment are $\epsilon_d-\mu$, $V_{sd}$ and $U$
(see Fig.~1). For convenience the calculations in this paper will be expressed through the parameters $\mu_L$, $\mu_R$ and $U$ (or $\mu$, $V_{sd}$ and $U$). However, one can easily solve Eq.~\ref{para_eq} to find the corresponding value of $\mu$ for a given $\epsilon_d-\mu$. Obviously $\mu=0$ (or $\mu_R=
-\mu_L=V_{sd}/2$) corresponds to the particle-hole symmetric point $\epsilon_d-\mu=-U/2$ (see Fig.~1).
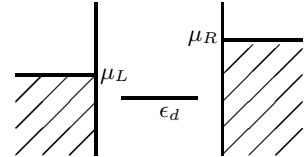
\begin{figure}
\begin{center}
\setlength{\unitlength}{.1in}
\begin{picture}(15,11)(0,0)
\linethickness{1pt}
\put(4.4,4){$\mu_L$}
\put(4.2,0){\line(0,1){8}}
\put(0,4.2){\line(1,0){4.2}}
\put(4.2,4.2){\line(-1,-1){4.2}}
\put(2.8,4.2){\line(-1,-1){2.8}}
\put(1.4,4.2){\line(-1,-1){1.4}}
\put(4.2,2.8){\line(-1,-1){2.8}}
\put(4.2,1.4){\line(-1,-1){1.4}}

\put(9,6){$\mu_R$}
\put(10.8,0){\line(0,1){8}}
\put(10.8,0){\line(1,1){4.2}}
\put(12.2,0){\line(1,1){2.8}}
\put(13.6,0){\line(1,1){1.4}}
\put(10.8,1.4){\line(1,1){4.2}}
\put(10.8,2.8){\line(1,1){3.0}}
\put(10.8,4.2){\line(1,1){1.6}}
\put(10.8,6){\line(1,0){4.2}}

\put(5.5,3){\line(1,0){4}}
\put(7.5,2){$\epsilon_d$}
\end{picture}
\label{fig1}
\caption{Schematic representation of the parameters in the Anderson impurity model.}
\end{center}
\end{figure}

The flow equation approach employs suitable infinitesimal unitary transformations in order
to diagonalize a given many-particle Hamiltonian. Thereby a one parameter family $H(B)$ of
unitarily equivalent Hamiltonians is generated, where $H(B=0)$ is the initial Hamiltonian 
(\ref{model2}) and $H(B=\infty)$ the final diagonal Hamiltonian. Such a unitary flow can
be generated as the solution of the following differential equation
\begin{eqnarray}
\frac{dH(B)}{dB}=[\eta(B),H(B)] \ ,
\end{eqnarray}
where $\eta(B)$ is an anti-hermitean operator. Wegner showed \cite{wegner1994} that the so-called canonical
choice $\eta(B)=[H(B),H_{\rm int}(B)]$, where $H_{\rm int}(B)$ the interaction part of the Hamiltonian,
leads to the required renormalization group-like diagonalization scheme. Our key
approximation will be the restriction to second order in~$U$. In this approximation 
the generator $\eta(B)=\eta^{(1)}(B)+\eta^{(2)}(B)$
takes the following form (for more details see Ref.~\cite{Moeckel}):
\begin{widetext}
\begin{eqnarray}\nonumber
\eta^{(1)}(B)&=& 
\sum_{s'_1s'_2s_1s_2} (\epsilon_{s_1'} + \epsilon_{s_2'} - \epsilon_{s_1} -\epsilon_{s_2} ) 
 U B_{s_1'}B_{s_1}B_{s_2'}B_{s_2} 
 e ^ {-B (\epsilon_{s_1'} + \epsilon_{s_2'} - \epsilon_{s_1} -\epsilon_{s_2} )^2} 
:c^\dag_{s_1' \uparrow} c_{s_1 \uparrow} c^\dag_{s_2' \downarrow} c_{s_2 \downarrow}:,\\
\nonumber \eta^{(2)}(B)&=& U^2 \sum_{s'\neq s,s'_1s_2s'_2\sigma}  
\frac{B_{s'}B_sB^2_{s'_1}B^2_{s_2}B^2_{s'_2}}{\epsilon_{s'}-\epsilon_s}
 Q_{s_1's_2s_2'} 
e^{-B(\epsilon_{s'}+\epsilon_{s_2}-\epsilon_{s'_1}-\epsilon_{s'_2})^2-
B(\epsilon_{s}+\epsilon_{s_2}-\epsilon_{s'_1}-\epsilon_{s'_2})^2} \\
&& \times (\epsilon_{s'}+\epsilon_s+2\epsilon_{s_2}-2\epsilon_{s'_1}-2\epsilon_{s'_2})
:c^\dag_{s' \sigma}c_{s\sigma}:
\end{eqnarray}
\end{widetext}
where
\begin{eqnarray}
Q_{s_1's_2s_2'} \stackrel{\rm def}{=} n_{s_1'}n_{s_2'} - n_{s_1'}n_{s_2} + n_{s_2}(1-n_{s_2'}).
\end{eqnarray}
The flow of the single-particle energies plays no role in the thermodynamic limit if
one is interested in impurity correlation functions or the current.
Therefore the final diagonal Hamiltonian takes the following simple form
\begin{eqnarray}\label{hamiltonian_infty}
 H(B=\infty)= \sum_{k\sigma} \epsilon_k  c^\dag_{k - \sigma} c_{k - \sigma} + \sum_{s\sigma} \epsilon_s c^\dag_{s \sigma} c_{s \sigma}.
\end{eqnarray}
Here one should notice that energy-diagonal terms like 
$\delta_{\epsilon_{s'_1}+\epsilon_{s'_2},\epsilon_{s_1}+\epsilon_{s_2}} U B_{s_1'}B_{s_1}B_{s_2'}B_{s_2} :c^\dag_{s_1' \uparrow} c_{s_1 \uparrow} c^\dag_{s_2' \downarrow} c_{s_2 \downarrow}:$ still remain
in $H(B=\infty)$ but have been neglected in (\ref{hamiltonian_infty}). This is permitted since
these terms are thermodynamically irrelevant, that is they vanish  in the thermodynamic limit.

\section{Flow of the Current Operator}

Clearly, the time evolution generated by (\ref{hamiltonian_infty}) in the diagonal
basis is trivial. However, the price we have to pay is to express the observable we
are interested in in this diagonal basis \cite{hackl}. Specifically, we look at the current operator
$I= I_\uparrow + I_\downarrow$, where
\begin{eqnarray}\nonumber
 I_\sigma && =(\partial_t N_{L \sigma} - \partial_t N_{R \sigma} ) /2 \\
&& \nonumber =\frac{iV}{2} \sum_k (d^\dag_ \sigma c_{k - \sigma} - h.c.) \\
&& =\frac{iV}{2} \sum_{s,k} B_s (c^\dag_ {s \sigma} c_{k - \sigma} - h.c.).
\label{defcurrentop}
\end{eqnarray}
Due to spin symmetry we only need to calculate the spin-up current $I_\uparrow$ since $I_\uparrow(t)=I_\downarrow(t)$.

The Hamiltonian has been diagonalized by the unitary transformation $U(B)$ corresponding to the
generator $\eta(B)$ given above. We perform the same unitary transformation on the current operator 
\begin{eqnarray}\label{current_flow}
 \frac{dI_\uparrow(B)}{dB} = [\eta(B), I_\uparrow(B)]
\end{eqnarray}
with the initial condition that $I_\uparrow(B=0)$ is given by (\ref{defcurrentop}).
In the current operator the anti-symmetric combinations $c_{k-\uparrow}$ stay invariant under the unitary transformation, 
while the commutator of $c^\dag_{s\uparrow}$ and $\eta$
generates higher order terms like $:c^\dag_{s_1' \uparrow} c^\dag_{s_2' \downarrow} c_{s_2 \downarrow}:$.
The commutator between this term and $\eta$ feeds back into the coefficient of $c^\dag_{s\uparrow}$.
For the lowest order correction with interaction (second order in $U$), the ansatz of the flowing current operator looks like
 \begin{eqnarray}\label{current_ansatz}\nonumber
 I_\uparrow (B) &=& \sum_{sk} \gamma_s (B)\,c^\dag_ {s \uparrow} c_{k - \uparrow} 
\\ && \nonumber + \sum_{s_1' s_2' s_2 k} M_{\uparrow \downarrow \downarrow}^{s_1' s_2' s_2} (B) \,
 :c^\dag _ {s_1' \uparrow} c^\dag_{s_2' \downarrow} c_{s_2 \downarrow}: c_{k - \uparrow} \\&& + h.c..
 \end{eqnarray} 
Substituting this ansatz into Eq.~(\ref{current_flow}) one finds the following flow of parameters:
\begin{widetext}
 \begin{eqnarray}\label{flow_parameter}
\nonumber \partial_B \gamma_s &=& U \sum_{s_1' s_2' s_2} M_{\uparrow \downarrow \downarrow}^{s_1' s_2' s_2} Q_{s_1's_2s_2'} 
(\epsilon_s+\epsilon_{s_2}-\epsilon_{s_1'}-\epsilon_{s_2'})
 B_{s}B_{s_1'}B_{s_2}B_{s_2'} 
 e ^ {-B (\epsilon_{s} + \epsilon_{s_2} - \epsilon_{s_1'} -\epsilon_{s_2'} )^2}
 + U^2\sum_{s'\neq s,s_1's_2s_2'} \gamma_{s'}Q_{s_1's_2s_2'} 
  \\
 && \nonumber \times 
2(\frac{\epsilon_s + \epsilon_{s'}}{2} + \epsilon_{s_2} 
 - \epsilon_{s_1'} - \epsilon_{s_2'}) 
\frac{B_{s_1'}^2 B_{s_2}^2 B_{s_2 '}^2 B_sB_{s'}}{\epsilon_s - \epsilon_{s'}}
e ^{-B[( \epsilon_{s} + \epsilon_{s_2} - \epsilon_{s_1'}- \epsilon_{s_2'} )^2 + ( \epsilon_{s'} + \epsilon_{s_2} - \epsilon_{s_1'}- \epsilon_{s_2'} )^2  ]} \\
 \partial_B M_{\uparrow \downarrow \downarrow}^{s_1' s_2' s_2} &=& U\sum_{s_1} \gamma_{s_1} 
(\epsilon_{s_1'} + \epsilon_{s_2'} - \epsilon_{s_1} -\epsilon_{s_2} ) 
B_{s_1'}B_{s_1}B_{s_2'}B_{s_2} 
 e ^ {-B (\epsilon_{s_1'} + \epsilon_{s_2'} - \epsilon_{s_1} -\epsilon_{s_2} )^2}, 
 \end{eqnarray}
 \end{widetext}
The higher order term in $\partial_B M_{\uparrow \downarrow \downarrow}^{s_1' s_2' s_2}$ is neglected since
we take only terms up to second order in~$U$ into account.

Next we use the simple time evolution in the diagonal basis
\begin{eqnarray}\label{current_t}
I_\uparrow(B=\infty,t)= e^{iH(\infty)t} I_\uparrow(B=\infty) e^{-iH(\infty)t}
\end{eqnarray}
leading to
 \begin{eqnarray}\label{parameter_time}
 \nonumber \gamma_s(\infty,t) &=& \gamma_s(\infty) e^{it \epsilon_s}, \\
 M_{\uparrow \downarrow \downarrow}^{s_1' s_2' s_2} (\infty,t) &=& M_{\uparrow \downarrow \downarrow}^{s_1' s_2' s_2} (\infty)
 e^{it (\epsilon_{s_1'} + \epsilon_{s_2'} - \epsilon_{s_2})}.
 \end{eqnarray}
Next we undo the unitary transformation, that is we integrate (\ref{current_flow})
from $B=\infty$ with initial conditions (\ref{parameter_time}) to $B=0$:
\begin{eqnarray}\nonumber
 I_\uparrow (0,t) &=& \sum_{sk} \gamma_s (0,t)\,c^\dag_ {s \uparrow} c_{k - \uparrow} 
\\&& \nonumber + \sum_{s_1' s_2' s_2 k} M_{\uparrow \downarrow \downarrow}^{s_1' s_2' s_2} (0,t) \,  :c^\dag _ {s_1' \uparrow} c^\dag_{s_2' \downarrow} c_{s_2 \downarrow}: c_{k - \uparrow} \\&& + h.c..
 \end{eqnarray} 
Our target is actually $\gamma_s(0,t)$ in this expression as we will find in the next chapter that 
only $\gamma_s(0,t)$ contributes to the expectation value of the current.

The solution of Eq.~(\ref{flow_parameter}) to the second order in $U$ can be written as (see Appendix A)
\begin{eqnarray}\label{gamma_s}\nonumber
\gamma_s(0,t) &=& \frac{iVB_s}{2} e^{i\epsilon_s t} + 
\frac {iVB_s U^2}{2} \sum_{s_1,D} T(D) B_{s_1}^2  \\ && \nonumber \times
\left[ \frac{ e^{iDt}-e^{i\epsilon_s t}}{(\epsilon_s-D)(\epsilon_{s_1}-D)}
+\frac{e^{i\epsilon_s t}- e^{i\epsilon_{s_1} t}}{(\epsilon_s-\epsilon_{s_1})(\epsilon_{s_1}-D)}\right], \\
\end{eqnarray}
where 
\begin{eqnarray}\label{def_t}\nonumber
T(D) &=& \sum_{s_1' s_2'} Q_{s_1' (\epsilon_{s_1 '} + \epsilon_{s_2 '}
-D )s_2'}B_{s_1'}^2 B_{s_2'}^2 \\
&& \times B^2(\epsilon_{s_1 '} + \epsilon_{s_2 '}-D).
\end{eqnarray}

\section{Calculation of the current}
\label{sec:current}

At time $t=0$ the coupling between the leads and the impurity is switched on.
The initial state is the non-interacting ground state, so the expectation value
of the current operator can be obtained easily: The quartic term in Eq.~(\ref{current_ansatz})
is normal-ordered and does therefore not contribute to the expectation value. The time-dependent current
is expressed as
\begin{eqnarray}\label{current_non}
\nonumber I_\uparrow(t) &=& <I_\uparrow(0,t)>_0 \\ \nonumber
&=& \textbf{Re} \sum_{sk} \frac{\gamma_s(0,t) e^{-it \epsilon_k} VB_s}
{\epsilon_s - \epsilon_k}(f_{L}(\epsilon_k)-f_{R}(\epsilon_k))  . \\
\end{eqnarray}

With Eq.~(\ref{gamma_s}) this gives an explicit expression for the current (see Appendix B). The summation over $s_1$ and $s$ can be calculated analytically. However, one has to be careful since there
are poles in the function and the summation cannot be simply transformed 
into a principal value integration. We employ the following trick to circumvent this problem. For example,
when calculating $\sum_{s} \frac{B^2_s}{\epsilon_s-\epsilon_k} \frac{e^{iDt}- e^{i\epsilon_s t}}{\epsilon_s -D}$, we introduce a second time~$t'$ and write the expression as 
\begin{eqnarray}
f(t,t')= \sum_{s} \frac{B^2_s}{\epsilon_s-\epsilon_k}e^{iDt} \frac{1- e^{i(\epsilon_s-D) t'}}{\epsilon_s -D}.
\end{eqnarray}
Obviously $f(t,t)$ is the original function that we are interested in and $f(t,0)=0$.
Now the pole at $\epsilon_s=D$ can be eliminated by partial differentiation with respect to~$t'$: 
\begin{eqnarray}
\frac{\partial f}{\partial t'} = \sum_{s} \frac{B^2_s}{\epsilon_s-\epsilon_k}e^{iDt} (-i)e^{i(\epsilon_s-D) t'}. 
\end{eqnarray}
The poles at $\epsilon_s=\epsilon_k$ can be eliminated likewise (see details
in Ref.~\cite{pei}) and the result is $\sum_{s} \frac{B^2_s}{\epsilon_s-\epsilon_k}e^{i\epsilon_s t'} = \frac{e^{i\epsilon_k t'}-e^{-\Gamma t'}}{\epsilon_k-i\Gamma}$. Therefore
\begin{eqnarray}
\frac{\partial f}{\partial t'} =-i e^{iDt} \frac{e^{i(\epsilon_k-D) t'}-e^{-(iD+\Gamma)t'}}{\epsilon_k -i\Gamma}.
\end{eqnarray}
and the original function follows by integration, $f(t,t)= \int^t_0 dt' \frac{\partial f}{\partial t'}$.
The key idea of our method is to introduce the additional time parameter~$t'$ and to get rid of the poles by performing derivatives with respect to~$t'$. Afterwards one can convert the sum into an integral.
Finally one perform the integration with respect to~$t'$ and gets the original function.

We divide the current into the zeroth order term and interaction corrections (see Appendix B),
\begin{eqnarray}
 I(t) = I^{(0)}(t)+ I^{(c)}(t),
\end{eqnarray}
where
\begin{eqnarray}\label{current_zero} \nonumber
\frac{I^{(0)}(t)}{\Gamma /h} &=& 
\int d\epsilon(f_R(\epsilon)-f_L(\epsilon)) \\ && \times \left( \frac{2\Gamma}{\epsilon^2+\Gamma^2} +  2e^{-\Gamma t} \frac{\epsilon \sin \epsilon t -\Gamma \cos \epsilon  t}
{\epsilon^2+\Gamma^2}\right) 
\end{eqnarray}
and
\begin{equation}\label{current_correction} 
\begin{split} 
\frac{I^{(c)}(t)}{\Gamma /h} =& \int d\epsilon  (f_R(\epsilon)-f_L(\epsilon))  \frac{2U^2}{\Gamma} \int dD \tilde T(D) \\ 
&  \times \textbf{Re} \left[ \frac{ie^{i(\epsilon-D)t}-i}{
(D-\epsilon)(D+i\Gamma)^2}+ \frac{t e^{i\epsilon t- \Gamma t}}{(\epsilon+i\Gamma)(D+i\Gamma ) }  \right. \\ 
& \left.  + \frac{(e^{i\epsilon t-\Gamma t}-1)(iD+i\epsilon -2\Gamma)}{(\epsilon+i\Gamma)^2 (D+i\Gamma)^2} \right].
\end{split} 
\end{equation} 
The dimensionless function $\tilde T$ is defined as
\begin{eqnarray}\nonumber
\nonumber \tilde T(D) &=& \int d\epsilon_{s'_1}d\epsilon_{s'_2} \\ && \nonumber \times \frac { \Gamma^4 Q_{s'_1(\epsilon_{s'_1}+\epsilon_{s'_2}-D )s'_2}}
{\pi^3 (\Gamma^2+\epsilon^2_{s'_1})(\Gamma^2+\epsilon^2_{s'_2})[\Gamma^2+
(\epsilon_{s'_1}+\epsilon_{s'_2}-D)^2]}. \\
\end{eqnarray}
If one uses the hybridization $\Gamma$ as the unit  of energy and $1/\Gamma$ as the unit of time, 
one can write $\displaystyle \frac{I}{\Gamma/h}$ as a function of three dimensionless quantities:
$\tilde V_{sd}=V_{sd}/\Gamma$, $\tilde U=U/\Gamma $ and $\tilde t=\Gamma t $ with
\begin{eqnarray}
 \frac{I}{\Gamma/h} = I(\tilde t, \tilde V_{sd}, \tilde U).
\end{eqnarray}

Two limiting cases deserve special attention. First,
it is straightforward to verify that the current is actually zero at $t=0$ as required. 
The calculation of the steady state current when $t\to\infty$ is also not difficult. The terms proportional to $e^{-\Gamma t}$ vanish in this limit and we find after a short calculation:
\begin{equation}
\begin{split}
\lim_{t\to \infty} \frac{I(t)}{\Gamma/h} =& \int d\epsilon(f_R(\epsilon)-f_L(\epsilon)) \\ & \times\left( \frac{2\Gamma}{\epsilon^2+\Gamma^2} + \frac{4U^2 \epsilon}{(\epsilon^2+\Gamma^2)^2} \int dD \frac{\tilde T(D)}{\epsilon-D} \right. \\ & \left. + \frac{2\pi U^2}{\Gamma} \tilde T(\epsilon) \frac{\epsilon^2-\Gamma^2}
{(\epsilon^2+\Gamma^2)^2} \right) \ . 
\end{split}
\end{equation}

\section{Relation between the current and the impurity spectral density}

Using Green's function methods, the current can be expressed by the lesser Green's function as
\begin{eqnarray}\label{current_green}
 I_\uparrow(t)= \frac{V}{\sqrt{2}} \sum_k \textbf{Re} (G^<_{kL}(t,t)-G^<_{kR}(t,t)), 
\end{eqnarray}
where $G^<_{k\alpha}(t,t) = i \langle d^\dag_\uparrow (t) c_{k\alpha \uparrow}(t)\rangle_0$.
According to Meir and Wingreen \cite{meir}, the lesser Green's function is related to
the retarded impurity Green's function:
\begin{equation}
\begin{split}
 G^<_{k\alpha}(t,t)=&\int_0^\infty dt' \left( g^r_{k\alpha}(t,t')
\frac{V}{\sqrt{2}} G^<(t',t) \right. \\ & \left. + g^<_{k\alpha}(t,t')\frac{V}{\sqrt{2}} G^a(t',t) \right),
\end{split}
\end{equation}
where 
\begin{eqnarray}
 g^r_{k\alpha}(t,t')&=&-i\theta(t-t')e^{i\epsilon_k(t'-t)} \\
g^<_{k\alpha}(t,t')&=&ie^{i\epsilon_k(t'-t)} f_{k\alpha}
\end{eqnarray}
are the conduction band Green's functions and 
\begin{eqnarray}
 G^<(t,t') &=& i\langle d^\dag_\uparrow(t') d_\uparrow(t) \rangle_0 \\
G^a(t,t') &=& i \theta(t'-t) \langle \{d_\uparrow(t),d^\dag_\uparrow(t')\} \rangle_0.
\end{eqnarray}
are the impurity Green's functions.
Eq.~(\ref{current_green}) can therefore be rewritten
\begin{eqnarray}\nonumber \label{meir}
 I_\uparrow(t)&=& \frac{1}{2\pi} \int d\epsilon_k (f_{kL}-f_{kR}) \\
&& \times\textbf{Im} \int_0^\infty dt' e^{i\epsilon_k(t-t')} 
 G^r(t,t') \ ,
\end{eqnarray}
where we have used the relation $G^a(t,t')=G^{r*}(t',t)$. 

The retarded Green's function $G^r(t,t')$ defined above depends not only on the time difference $t-t'$. 
We therefore define a time-dependent impurity spectral density
\begin{eqnarray}
\rho(t,\epsilon) = \frac{-1}{\pi} \textbf{Im}G^r(t, \epsilon),
\end{eqnarray}
where $G^r(t,\epsilon)$ is defined via
\begin{eqnarray}
 G^r(t, \epsilon)= \int_0^\infty dt' e^{i \epsilon (t-t')} G^r(t,t') \ .
\end{eqnarray}
Now the time-dependent Meir-Wingreen formula relates the time-dependent current 
with the time-dependent impurity spectral density,
\begin{eqnarray}
I(t)= \int d\epsilon (f_R(\epsilon)-f_L(\epsilon)) \rho(t,\epsilon).
\end{eqnarray}

The flow equation result for the Heisenberg time evolution of $d^\dag_\sigma(t)$ 
has already been given in Sect.~\ref{sec:current}. 
Therefore the calculation of the time-dependent impurity spectral density is straightforward, details
can be found in Appendix~C.
Explicit comparison of Eqs.~(\ref{current_zero}) and (\ref{current_correction})
from the direct solution of the Heisenberg equations of motion for the current operator with 
Eq.~(\ref{spec_density}) shows that our previous results in Sect.~\ref{sec:current}
are consistent with the time-dependent Meir-Wingreen formula as should be expected. In the steady state
limit $t\to \infty$ we find the familiar equilibrium impurity spectral density
\begin{eqnarray}\nonumber
\lim_{t\to \infty} \rho(t,\epsilon) &=& \frac{\Gamma^2}{\pi(\epsilon^2+\Gamma^2)} + \frac{2U^2 \epsilon\Gamma}{\pi (\epsilon^2+\Gamma^2)^2} \int dD
\frac{\tilde T(D)}{\epsilon-D}  \\ 
&& + \frac{U^2 \tilde T(\epsilon)(\epsilon^2-\Gamma^2)}{(\epsilon^2+\Gamma^2)^2}. 
\end{eqnarray}
This equation reproduces the result in Ref.~\cite{ueda}. 

\section{Time-dependent current at particle-hole symmetry}

The above formulas for time-dependent current and spectral density hold for arbitrary left and right
lead chemical potentials. In the sequel we will present some explicit results for the time-dependent
current at the particle-hole symmetric point, $\epsilon_d- (\mu_L+\mu_R)/2 = -U/2$.

We perform numerical integration to get the time-dependent current curves. A direct estimation
of Eq.~(\ref{current_correction}) is difficult because there is a pole in the integrand. Alternatively, 
we calculate the time derivative of the current, i.e.
\begin{eqnarray} \label{current_corr_diff}
\frac{d}{dt}\left(\frac{I^{(c)}(t)}{\Gamma/h}\right)&=& \frac{4U^2 \sin \frac{V_{sd}}{2} t}{\Gamma t}
\int dD \: \tilde T(D) 
\\ && \times \left(\textbf{Re} \frac{e^{-iDt}-e^{-\Gamma t}}{(D+i\Gamma)^2} + \frac{\Gamma t e^{-\Gamma t}}{D^2+\Gamma^2}\right). \nonumber
\end{eqnarray}
We then perform numerical integration of the right side in (\ref{current_corr_diff}) and employ a 
fourth-order Runge-Kutta method to solve (\ref{current_corr_diff}) and 
get the current. The symmetry of $\tilde T$ function, i.e. $\tilde T(-D)=\tilde T(D)$, is used
to simplify the calculation. 

\begin{figure}
\begin{center}
\includegraphics[width=0.45\textwidth]{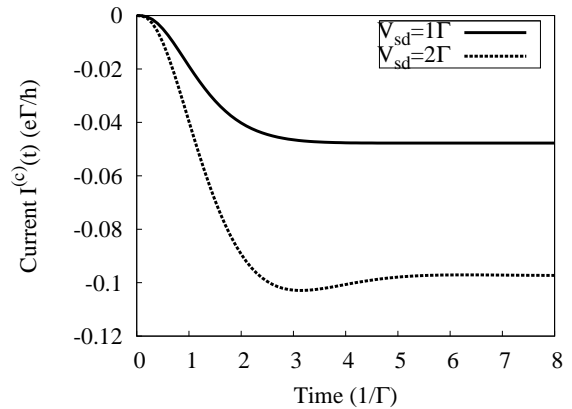}
\caption{The current correction $I^{(c)}(t)$ due to interaction at particle-hole symmetry,
$\epsilon_d= - U/2$, for zero temperature. The interaction strength is $U=\Gamma$. 
Results for voltage bias $V_{sd}=\Gamma$ and $V_{sd}=2\Gamma$
are depicted. The main features of $I^{(c)}(t)$ are a vanishing
derivative at $t=0$, followed by a sharp decrease and finally a smooth crossover
towards its steady value. One also notices the onset of oscillations at large voltage bias $V_{sd}=2\Gamma$. }
\end{center}
\end{figure}

\begin{figure}
\begin{center}
\includegraphics[width=0.45\textwidth]{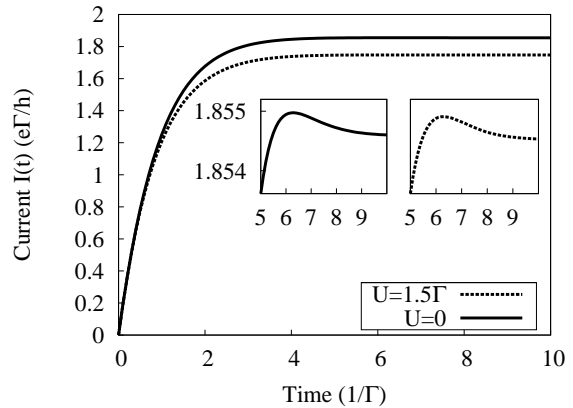}
\caption{The current without interaction and for interaction strength $U=1.5\Gamma$
at voltage bias $V_{sd}=\Gamma$. The interaction suppresses the current. The inset
shows the suppressed oscillation of the current.}
\end{center}
\end{figure} 
\begin{figure}
\begin{center}
\includegraphics[width=0.45\textwidth]{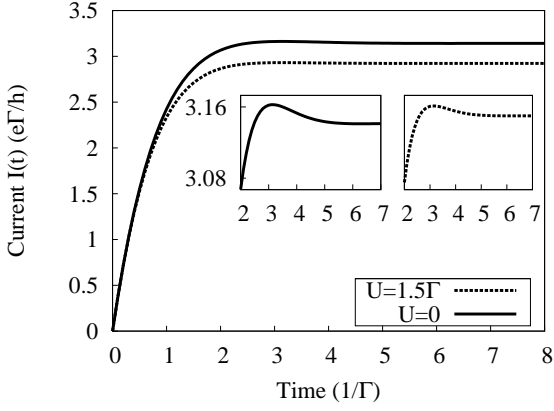}
\caption{The current without interaction and for interaction strength $U=1.5\Gamma$
at voltage bias $V_{sd}=2\Gamma$. The free current increases 
compared to $V_{sd}=\Gamma$, while its interaction suppression also becomes stronger
due to shot noise induced decoherence.
The inset shows suppressed current oscillation.}
\end{center}
\end{figure} 
Fig.~2 shows the interaction correction to the current at different voltage bias. Its time 
derivative at $t=0$ vanishes. This is contrary to the free current, which has a
sharp increase at $t=0$ (see Figs.~3,~4), which indicates the initial condition 
$n_d=0$. However, this initial charging process is independent of~$U$ 
due to the lack of electrons in the impurity, which explains $\frac{d}{dt}I^{(c)}(t=0)=0$.

For $t\gg 1/\Gamma$ the current correction approaches its steady value. Larger voltage bias leads to 
a stronger suppression of the current due to the $U^2$-dependent correction term.
This can be understood to arise from shot noise decoherence effects, which suppress the quasiparticle
resonance, similar to the well-established effect of current-induced decoherence
in the nonequilibrium Kondo model \cite{Rosch2001}.

The suppressed ringing oscillation in both current correction and total current
can be seen at large voltage bias $V_{sd}=2\Gamma$ (see Figs.~3 and~4). From
(\ref{current_zero}) and (\ref{current_corr_diff}) one can easily deduce the ringing oscillation period $4\pi/V_{sd}$, consistent with Ref.~\cite{wingreen94}.

\section{Conclusions}
We have demonstrated how the flow equation method (method of infinitesimal unitary
transformations) can be used to calculate transient and steady state currents 
in and beyond the linear response regime
through interacting quantum impurities. Our approach is perturbative in nature,
therefore we are restricted to weak to intermediate values of the interaction
in our analysis of the Anderson impurity model in this paper. One key feature
of our approach is that there are no secular terms in the long time limit, 
that is the steady state is reached uniformly in the expansion in the interaction.
We reproduce previous results for the steady state currents \cite{ueda} and
obtain analytical results for the transient current behavior leading to the
steady state.

\begin{acknowledgments}
We thank M.~Moeckel for valuable discussions.
We acknowledge support through SFB~484 of the Deutsche Forschungsgemeinschaft,
the Center for NanoScience (CeNS) Munich, and the German Excellence Initiative via
the Nanosystems Initiative Munich (NIM). 
\end{acknowledgments}

\appendix

\section{Solution for $\gamma_s(0,t)$}

The differential equation~(\ref{flow_parameter}) is solved order by order in~$U$.
According to the definition of the current operator, we have the initial condition $\gamma_s(0,0)=\frac{iV}{2} B_s$
and $M(0,0)= 0$.
The zeroth order solution can be written as $M(B,t)=0$ and $\gamma_s(B,t)=\frac{iV}{2} B_s e^{it \epsilon_s} $ according to Eq.~(\ref{parameter_time}).
Substituting $\gamma_s(B,t)$ into Eq.~(\ref{flow_parameter}) and integrating with respect to $B$ at $t=0$,
we get
\begin{eqnarray}\nonumber
 M_{\uparrow \downarrow \downarrow}^{s_1' s_2' s_2 } (B,0) &=& 
 iVU \sum_{\epsilon_{s_1}\neq \epsilon_{s'_1}+\epsilon_{s'_2}-\epsilon_{s_2}}  B_{s_1'}B_{s_1}^2 B_{s_2'}B_{s_2} \\
&& \times \frac{1 - 
 e^{-B (\epsilon_{s_1'}+ \epsilon_{s_2'}-\epsilon_{s_1}-\epsilon_{s_2})^2 } }
 {2(\epsilon_{s_1'}+ \epsilon_{s_2'}-\epsilon_{s_1}-\epsilon_{s_2})}. 
\end{eqnarray}
Integrating with respect to $B$ for a given time~$t$ one finds the first order solution of $M$,
 \begin{equation}\label{M_expr}
 \begin{split}
 M_{\uparrow \downarrow \downarrow}^{s_1' s_2' s_2 } (B,t) =& 
 iVU \sum_{s_1}  B_{s_1'}B_{s_1}^2 B_{s_2'}B_{s_2} \\& \times \left(
\frac{e^{it(\epsilon_{s_1'} + \epsilon_{s_2'} - \epsilon_{s_2})}  }
 {2(\epsilon_{s_1'}+ \epsilon_{s_2'}-\epsilon_{s_1}-\epsilon_{s_2})} \right. \\ &
 \left. - \frac{ e^{it \epsilon_{s_1}-B (\epsilon_{s_1'}+ \epsilon_{s_2'}-\epsilon_{s_1}-\epsilon_{s_2})^2 }}{ 2(\epsilon_{s_1'}+ \epsilon_{s_2'}-\epsilon_{s_1}-\epsilon_{s_2})} \right).
 \end{split}
 \end{equation}
Taking the limit $B\to \infty$ we find
\begin{eqnarray}\nonumber
 M_{\uparrow \downarrow \downarrow}^{s_1' s_2' s_2 } (\infty,t)
=iVU \sum_{s_1}  
 \frac{B_{s_1'}B_{s_1}^2 B_{s_2'}B_{s_2} e^{it(\epsilon_{s_1'} + \epsilon_{s_2'} - \epsilon_{s_2})}}
 {2(\epsilon_{s_1'}+ \epsilon_{s_2'}-\epsilon_{s_1}-\epsilon_{s_2})}.\\
\end{eqnarray}
Substituting the above expression and the zeroth order solution of $\gamma_s$ into Eq.~(\ref{flow_parameter}),
we find the solution of $\gamma_s$ to second order in~$U$,
\begin{equation}
\begin{split}
\delta \gamma_s(t) =& \gamma_s(\infty,t)- \gamma_s(0,t) \\ 
=& \frac {iVB_s U^2}{2} \sum_{s_1,D} T(D) B_{s_1}^2 
\left[ \frac{ -e^{itD}}{(\epsilon_s-D)(\epsilon_{s_1}-D)} \right. \\& \left.
+\frac{ e^{it\epsilon_{s_1}}}{(\epsilon_s-\epsilon_{s_1})(\epsilon_{s_1}-D)} \right],
\end{split}
\end{equation}
 where $D= \epsilon_{s_1 '} + \epsilon_{s_2 '} - \epsilon_{s_2} $ and $T(D)$ is defined in Eq.~(\ref{def_t}).
Then we have
\begin{equation}
\begin{split}
\gamma_s(0,t) =& e^{i\epsilon_s t} (\gamma_s(0,0) +
\delta\gamma_s(0)) - \delta \gamma_s(t) \\ =& \frac{iVB_s}{2} e^{i\epsilon_s t} + 
\frac {iVB_s U^2}{2} \sum_{s_1,D} T(D) B_{s_1}^2 \\& \times
\left[ \frac{ e^{iDt}-e^{i\epsilon_s t}}{(\epsilon_s-D)(\epsilon_{s_1}-D)}
+\frac{e^{i\epsilon_s t}- e^{i\epsilon_{s_1} t}}{(\epsilon_s-\epsilon_{s_1})(\epsilon_{s_1}-D)}\right].
\end{split}
\end{equation}

\section{The calculation of the current}

We divide the expression of the current into the zeroth order term and the interaction correction, $ I_\uparrow(t) = I_\uparrow^{(0)}(t) + I_\uparrow^{(c)}(t)$, where
\begin{eqnarray}\nonumber
I_\uparrow^{(0)}(t)
&=& \textbf{Re} \sum_{s,k} \frac{iV^2B^2_s}{2(\epsilon_s-\epsilon_k)} e^{i(\epsilon_s-\epsilon_k) t} (f_{L}(\epsilon_k)-f_{R}(\epsilon_k)),\\
\end{eqnarray}
and
\begin{equation}\label{current_c_proc}
 \begin{split}
  I_\uparrow^{(c)}(t)=& \textbf{Re}
\sum_{s,k,s_1,D} \frac {iV^2B^2_s U^2 e^{-i\epsilon_k t}}{2(\epsilon_s-\epsilon_k)}  T(D) B_{s_1}^2 \\ & \times
\left[ \frac{ e^{iDt}-e^{i\epsilon_s t}}{(\epsilon_s-D)(\epsilon_{s_1}-D)}
+\frac{e^{i\epsilon_s t}- e^{i\epsilon_{s_1} t}}{(\epsilon_s-\epsilon_{s_1})(\epsilon_{s_1}-D)}\right] \\&\times (f_{L}(\epsilon_k)-f_{R}(\epsilon_k)) . 
\end{split}
\end{equation}
The sum over $s$ and $s_1$ is calculated analytically by the method introduced
in Section IV. The sum over $s$ in the zeroth order term is straightforward.
Next we need to calculate
\begin{equation}
\begin{split}
\Lambda =& \sum_{s,s_1} \frac{B^2_s}{\epsilon_s-\epsilon_k} B^2_{s_1} \left[\frac{e^{iDt}-e^{i\epsilon_s t}}
{(\epsilon_s-D)(\epsilon_{s_1}-D)} \right. \\ &\left. + \frac{e^{i\epsilon_s t}-e^{i\epsilon_{s_1} t}}
{(\epsilon_s-\epsilon_{s_1})(\epsilon_{s_1}-D)} \right].
\end{split}
\end{equation}
We first calculate the sum over $s$ and get
\begin{equation}
\begin{split}
\Lambda =& \frac{1}{\epsilon_k-i\Gamma} \sum_{s_1}\frac{B^2_{s_1}}{\epsilon_{s_1}-D} \left[\frac{e^{i\epsilon_k t}-e^{iDt}}{D-\epsilon_k} + \frac{e^{iDt}-e^{-\Gamma t}}{D-i\Gamma} \right. \\& \left. + \frac{e^{i\epsilon_{s_1}t}-e^{i\epsilon_kt}}{\epsilon_{s_1}-\epsilon_k}+\frac{e^{-\Gamma t}-e^{i\epsilon_{s_1} t}}{\epsilon_{s_1}-i\Gamma} \right] .
\end{split}
\end{equation}
When calculating the sum over $s_1$, we have to get rid of the poles at $\epsilon_{s_1}=D$. We rearrange the terms so that $\epsilon_{s_1}-D$ in the denominator and $e^{i\epsilon_{s_1}t}- e^{iDt}$ in the numerator appear simultaneously, i.e.
\begin{equation}
\begin{split}
\Lambda =&\frac{1}{\epsilon_k-i\Gamma} \left[ \frac{i(e^{iDt}-e^{-\Gamma t})}{2\Gamma(D-i\Gamma)}+
\sum_{s_1} B^2_{s_1}\frac{e^{i\epsilon_{s_1}t}-e^{iDt}}{(\epsilon_{s_1}-D)(\epsilon_{s_1}-\epsilon_k)} \right. \\ & \left. + \sum_{s_1} B^2_{s_1}\frac{e^{iDt}-e^{i\epsilon_{s_1}t}}{(\epsilon_{s_1}-D)(\epsilon_{s_1}-i\Gamma)} \right] .
\end{split}
\end{equation}
Employing the method from Section~IV again we find
\begin{equation}
\begin{split}
\Lambda  =& \frac{1}{\epsilon_k-i\Gamma} \left[ \frac{1}{\epsilon_k-i\Gamma} (\frac{e^{i\epsilon_k t}-e^{iDt}}{\epsilon_k-D}+  \frac{e^{-\Gamma t}-e^{iDt}}{D-i\Gamma}) \right. \\ & \left.+ \frac{e^{-\Gamma t}-e^{iDt}}{(D-i\Gamma)^2} + \frac{it e^{-\Gamma t}}{D-i\Gamma}\right]  .
\end{split}
\end{equation}
Substituting the expression for $\Lambda$ into (\ref{current_c_proc}) we obtain an
expression for $I^{(c)}$.
The pole at $\epsilon_k =D$ is a removable singularity, so that we can change the sum over $k$ and $D$ into a Cauchy principal value integral. This transformation makes it easy to estimate the
long time limit and to compare our result with that in Ref.~\cite{ueda}. The interaction
correction for the current is then given by
\begin{equation}
\begin{split}
I_\uparrow^{(c)}(t) 
=& \int dD d\epsilon \frac{U^2 \tilde T(D)}{2\pi} (f_R(\epsilon)-f_L(\epsilon)) \\ & \times \textbf{Re} \left[\frac{ie^{i(\epsilon-D)t}-i}{
(D-\epsilon)(D+i)^2} + \frac{(e^{i\epsilon t-t}-1)(iD+i\epsilon -2)}{(\epsilon+i)^2 (D+i)^2} \right. \\& \left.
+ \frac{te^{i\epsilon t-t}}{(\epsilon+i)(D+i) }\right] .
\end{split}
\end{equation}

\section{The calculation of the spectral density}

The evolution of the $d^\dag_\sigma$ operator is similar to the current operator and can be expressed as
\begin{eqnarray}\nonumber
d^\dag_\uparrow(t) &=& \sum_s \tilde \gamma_s(0,t) c^\dag_{s\uparrow}
\\&& + \sum_{s'_1s'_2 s_2} \tilde M^{s'_1 s'_2 s_2}_{\uparrow \downarrow \downarrow}
(0,t) : c^\dag_{s'_1\uparrow}c^\dag_{s'_2\downarrow}c_{s_2\downarrow}: , 
\end{eqnarray}
where $ \tilde \gamma_s(0,t) = \frac{2}{iV} \gamma_s(0,t) $ and $ \tilde M^{s'_1 s'_2 s_2}_{\uparrow \downarrow \downarrow} (0,t) =   \frac{2}{iV} M^{s'_1 s'_2 s_2}_{\uparrow \downarrow \downarrow} (0,t)$. The anticommutator is
\begin{eqnarray}\nonumber
\langle \{ d(t), d^\dag(t') \}\rangle &=& \sum_s \tilde \gamma^*_s(0,t) \tilde \gamma_s(0,t')
\\ && \nonumber+ \sum_{s'_1s'_2s_2} \tilde M^{s'_1 s'_2 s_2 *}_{\uparrow \downarrow \downarrow} (0,t) \tilde M^{s'_1 s'_2 s_2}_{\uparrow \downarrow \downarrow} (0,t')\\ && \times Q_{s'_1s_2s'_2} .
\end{eqnarray}
By using the summation method from the calculation of the current, we find
\begin{equation}
\begin{split}
\sum_s &\tilde \gamma^*_s(0,t) \tilde \gamma_s(0,t') \\=& e^{\Gamma (t'-t)} + U^2T(D)\\& \times \left[
\frac{2\Gamma(t'-t)e^{\Gamma(t'-t)}}{2i\Gamma(D+i\Gamma)} + \frac{e^{-iDt+iDt'}-e^{\Gamma(t'-t)}}{(D+i\Gamma)^2}\right. \\& \left. +\frac{e^{-iDt-\Gamma t'}-e^{iDt'-iDt}
+ e^{iDt'-\Gamma t}- e^{-\Gamma (t+t')}}{D^2+\Gamma^2}  \right].
\end{split}
\end{equation}
Setting $B=0$ and performing the summation over $s_1$ in Eq.~\ref{M_expr}, we get
\begin{eqnarray}
\tilde M^{s'_1 s'_2 s_2}_{\uparrow \downarrow \downarrow}
(0,t) = UB_{s'_1}B_{s'_2}B_{s_2} \frac{e^{-\Gamma t}-e^{it(\epsilon_{s'_1}+\epsilon_{s'_2}
-\epsilon_{s_2})}}{\epsilon_{s_2}-\epsilon_{s'_1}-\epsilon_{s'_2}+i\Gamma}.
\end{eqnarray}
Using the definition $D=\epsilon_{s'_1}+\epsilon_{s'_2}-\epsilon_{s_2}$, we obtain
\begin{equation}
\begin{split}
\langle \{ d(t), d^\dag(t') \}\rangle =&e^{\Gamma (t'-t)} + U^2T(D)\\& \times \left[
\frac{2\Gamma(t'-t)e^{\Gamma(t'-t)}}{2i\Gamma(D+i\Gamma)} \right. \\& \left.+\frac{e^{-iDt+iDt'}-e^{\Gamma(t'-t)}}{(D+i\Gamma)^2} \right] .
\end{split}
\end{equation}
The impurity orbital spectral density is therefore given by
\begin{equation} \label{spec_density}
\begin{split}
\rho(t,\epsilon)=&\frac{1}{\pi(\epsilon^2+1)}+ \frac{e^{-\Gamma t}(\epsilon \sin \epsilon t-\cos \epsilon t)}{\pi(\epsilon^2+1)} \\ & + \textbf{Re} \frac{U^2 \tilde T(D)}{\pi} \left[ \frac{ie^{i(\epsilon-D)t}-i}{(D-\epsilon)(D+i)^2} + \frac{te^{i\epsilon t-\Gamma t}}{(D+i)(\epsilon+i)} \right. \\& \left. + \frac{(e^{i\epsilon t-\Gamma t}-1)(iD+i\epsilon-2)}{(D+i)^2(\epsilon+i)^2} \right].
\end{split}
\end{equation}

\end{document}